\documentclass[aps,prl,superscriptaddress,twocolumn,showpacs]{revtex4}

\usepackage{bm}
\usepackage{graphicx}
\usepackage{amsfonts}
\usepackage{amsmath}

\begin{document}

\title{Entanglement and factorized ground states in two-dimensional quantum 
antiferromagnets}

\author{Tommaso Roscilde}
%\email{roscilde@usc.edu}
\affiliation{Department of Physics and Astronomy, University of 
Southern California, Los Angeles, CA 90089-0484}
\author{Paola Verrucchi}
%\email{verrucchi@fi.infn.it}
\affiliation{\mbox{Istituto Nazionale per la Fisica della Materia, UdR Firenze,
Via G. Sansone 1, I-50019 Sesto F.no (FI), Italy}}
\affiliation{\mbox{Istituto Sistemi Complessi - C.N.R., 
Sez. di Firenze, via Madonna del Piano, I-50019 Sesto F.no (FI), Italy}}    
\author{Andrea Fubini}
%\email{fubini@fi.infn.it}
\affiliation{\mbox{Istituto Nazionale per la Fisica della Materia, UdR Firenze,
Via G. Sansone 1, I-50019 Sesto F.no (FI), Italy}}
\affiliation{\mbox{Dipartimento di Fisica dell'Universit\`a di Firenze,
Via G. Sansone 1, I-50019 Sesto F.no (FI), Italy}}  
\author{Stephan Haas}
%\email{roscilde@usc.edu}
\affiliation{Department of Physics and Astronomy, University of 
Southern California, Los Angeles, CA 90089-0484}    
\author{Valerio Tognetti}
\affiliation{\mbox{Istituto Nazionale per la Fisica della Materia, UdR Firenze,
Via G. Sansone 1, I-50019 Sesto F.no (FI), Italy}}
\affiliation{\mbox{Dipartimento di Fisica dell'Universit\`a di Firenze,
Via G. Sansone 1, I-50019 Sesto F.no (FI), Italy}}  
\affiliation{\mbox{Istituto Nazionale di Fisica Nucleare, Sez.
di Firenze, Via G. Sansone 1, I-50019 Sesto F.no
(FI), Italy}}

\date{\today}

\begin{abstract}

Making use of exact results and quantum Monte Carlo data for the entanglement
of formation, we show that the ground state of anisotropic two-dimensional
$S{=}1/2$ antiferromagnets in a uniform field takes the classical-like
form of a product state for a particular value and orientation of the field, at
which the purely quantum correlations due to entanglement disappear.  
Analytical expressions for the energy and the form of such states are given,
and a novel type of exactly solvable two-dimensional quantum models is
therefore singled out.  Moreover, we show that the field-induced quantum phase
transition present in the models is unambiguously characterized by a \emph{cusp
minimum} in the pairwise-to-global entanglement ratio $R$, marking the
quantum-critical enhancement of \emph{multipartite} entanglement. A detailed
discussion is provided on the universality of the cusp in $R$ as a signature of
quantum critical behavior entirely based on entanglement.

\end{abstract}

\pacs{03.67.Mn, 75.10.Jm, 73.43.Nq, 05.30.-d}
% 75.10.Jm - Quantized spin models
% 05.30.-d - Quantum statistical mechanics
% 73.43.Nq - Quantum phase transitions
% 03.67.Mn - Entanglement production, characterization, and manipulation
\maketitle

 The study of entanglement in quantum many-body systems
is an emerging field of research which promises to 
shed new light on our understanding of complex quantum 
models. Entanglement represents indeed a unique form
of correlation that quantum states do not share with
their classical counterparts, and it thus accounts for 
the purely quantum aspects of the many-body behavior. 
Given that collective phenomena
show up in a dramatic fashion at phase transitions,
the study of entanglement in quantum critical systems
represents an intriguing 
subject~\cite{Osterlohetal02,OsborneN02}.
The main focus has been so far the 
study of spin-$1/2$ chains, mainly because they provide 
paradigmatic examples of  exactly solvable 
quantum systems showing a quantum phase transition~\cite{BarouchMD71}. 
Only few studies~\cite{Syljuasen03-2D} have explored models in two 
dimensions and higher, and it is therefore hard to tell to which extent 
the behavior of entanglement observed at one-dimensional (1D) 
transitions reflects universal critical features~\cite{Wuetal04}.

A similarly intriguing question in this framework is: 
can we learn anything from entanglement that we did not know 
already from conventional equilibrium observables? 
In general entanglement at $T{=}0$ gives a unique insight on 
the global properties of the ground-state wavefunction,
measuring in a sense how far the quantum ground state sits 
from any possible classical counterpart. Therefore 
entanglement is a powerful tool to detect the occurrence,
concrete albeit surprising, of classical-like states 
in strongly interacting quantum systems.   

The aim of this letter is to show that a proper analysis of 
entanglement estimators, applied to interacting 
quantum spin systems in arbitrary dimensions, not only exhibits 
universal features at continuous quantum phase transitions, 
but also unveils the occurrence of unexpected ground-state 
features.

We consider the two-dimensional (2D) antiferromagnetic spin-$1/2$ XYZ 
model in a uniform magnetic field:
 \begin{equation}
  {\hat{\cal H}}/J =
   \sum_{\langle ij \rangle} \Big[ \hat{S}^x_i\hat{S}^x_{j}
   + \Delta_y \hat{S}^y_i\hat{S}^y_{j} 
   + \Delta_z \hat{S}^z_i\hat{S}^z_{j}\Big] - \sum_i {\bm h}\cdot\hat{\bm S}_i~,
  \label{e.XYZhz}
 \end{equation}
where $J{>}0$ is the exchange coupling, $\langle ij \rangle$ 
runs over the pairs of nearest neighbors, and 
${\bm h}{\equiv} g\mu_{\rm B} {\bm H}/J$ is the reduced magnetic field.
For the sake of simplicity we will hereafter understand the canonical 
transformation $\hat{S}^{x,y}_i {\to}({-}1)^{I} \hat{S}^{x,y}_i$ 
with $I=1(2)$ for $i$ belonging to sublattice $1(2)$, which
transforms the coupling in the $xy$ plane from antiferromagnetic
to ferromagnetic.
Eq.(\ref{e.XYZhz}) is the most general Hamiltonian for
an anisotropic spin-1/2 system with exchange
spin-spin interactions. However, 
as real compounds usually display axial symmetry,
we will henceforth consider either $\Delta_y{=}1$ or $\Delta_z{=}1$.
Moreover, we will apply the field along the $z$-axis, i.e. 
$\bm h{=}(0,0,h)$.
The case $\Delta_y{=}1$ corresponds to the XXZ model in a longitudinal
field, where no genuine quantum phase transition occurs upon 
changing the applied field.

 This paper focuses on the less investigated case
 $\Delta_z{=}1$, defining the XYX model in a field. 
 Due to the non-commutativity of the Zeeman and the exchange
 term, for $\Delta_y {\neq} 1$ this model is expected 
 to show a field-induced quantum
 phase transition on any D-dimensional bipartite lattice,
 with the universality class of the D-dimensional Ising model
 in a transverse field~\cite{Chakrabartietal96}. 
The two cases $\Delta_y{{<}}1$ and $\Delta_y{>}1$  
correspond to an easy-plane (EP) and easy-axis (EA) behavior,
respectively. 
The ordered phase in the EP(EA) case arises by spontaneous 
simmetry breaking along the $x$($y$) direction, which 
corresponds to a finite value of the order parameter 
$M^x$($M^y$) below the critical field $h_{\rm c}$.
At the transition, long-range correlations are
destroyed, and the system is left in a partially
 polarized state with field-induced magnetization
 reaching saturation only as $h{\to}\infty$. This picture 
 has been verified so 
 far in D=1 only, both analytically~\cite{Dmitrievetal02}
 and numerically~\cite{Cauxetal03,Roscildeetal04}.

We investigate the ground state properties of the XYX model
on a $L\times L$ square lattice making use of Stochastic Series 
Expansion quantum Monte Carlo simulations 
using a modified version of the directed-loop 
algorithm~\cite{SyljuasenS02,Roscildeetal04} to account for the low symmetry 
($Z_2$ only) of the Hamiltonian. With this numerical 
approach we can reach sizes as big as $L{=}28$ 
at very low temperatures ($T/J{=}1/2L$), such that 
the $T{=}0$ limit of the model, and in particular its
quantum-critical behavior, is accurately captured.

The main focus of our simulations is on 
entanglement properties, analysed through  
{\it one-tangle} and {\it concurrence}. 
The one-tangle quantifies the entanglement between
a spin and the remainder of the system. It is
defined as $\tau_1{=}4\det\rho^{(1)}$,
where $\rho^{(1)}$ is the one-site reduced density 
matrix~\cite{Coffmanetal00,Amicoetal04}. 
In terms of the magnetic observables, $\tau_1$ takes the simple form
$\tau_1{=}1{-}4\sum_\alpha (M^{\alpha})^2$, where $\alpha{=}x,y,z$, and
$M^{\alpha}{\equiv}\langle \hat{S}^{\alpha} \rangle$. 
The one-tangle represents a \emph{global}
 estimate of the entanglement in a translationally invariant
 system, since the vanishing of $\tau_1$ is a necessary
 and sufficient condition for the ground state to 
 be factorized.
 The concurrence~\cite{Wootters98} quantifies 
the pairwise entanglement between two spins at sites
$i$ and $j$. For the model of interest, in absence of
spontaneous simmetry breaking (SSB), the 
concurrence takes the form~\cite{Amicoetal04}
$C_{ij}{=}2~{\rm max}\{0,C_{ij}^{(1)},C_{ij}^{(2)}\}$,
where
\begin{eqnarray}
C_{ij}^{(1)}&=&g_{ij}^{zz}-{\textstyle\frac{1}{4}}+|g_{ij}^{xx}-g_{ij}^{yy}|~,
\label{e.C1}\\
C_{ij}^{(2)}&=&|g_{ij}^{xx}+g_{ij}^{yy}|-
[({\textstyle \frac{1}{4}}+g_{ij}^{zz})^2-(M^z)^2]^{1/2}~,
\label{e.C2}
\end{eqnarray}
with 
$g_{ij}^{\alpha\alpha}{\equiv}\langle\hat{S}_i^\alpha\hat{S}_{j}^\alpha\rangle$.
When SSB occurs, Sylju\aa sen~\cite{Syljuasen03} has shown 
that Eqs.~(\ref{e.C1})-(\ref{e.C2}) remain unchanged if the condition
$C_{ij}^{(2)} {<} C_{ij}^{(1)}$ is satisfied, otherwise
Eq.~(\ref{e.C2}) provides an upper bound for the actual concurrence.

One-tangle and  concurrence are related by the Coffman-Kundu-Wootters 
(CKW) conjecture~\cite{Coffmanetal00}
$\tau_1{\geq}\tau_2{\equiv}\sum_{j{\neq}i}C_{ij}^2$, which expresses
 the crucial fact that pairwise entanglement does not exhaust the global 
entanglement of the system, as entanglement can also be stored 
in $3$-spin correlations, $4$-spin correlations, and so on. The fact 
that $n$-spin entanglement
and $m$-spin entanglement with $m{\neq} n$ are mutually exclusive
is a unique feature of entanglement as a form of correlation,
which puts it at odds with classical correlations. 
In this respect, if the CKW conjecture is verified,
the \emph{entanglement ratio}~\cite{Roscildeetal04} 
$R{\equiv}\tau_2/\tau_1{<}1$ quantifies the relative weight of pairwise
entanglement, and its deviation from unity reveals in turn 
the relevance of 
$n$-spin entanglement with $n{>}2$. Although indirect,
the entanglement ratio is the only accessible 
estimate of multi-spin entanglement that we can systematically
implement at the moment. 
\begin{figure}
 \begin{center}
\includegraphics[bbllx=0pt,bblly=-40pt,bburx=550pt,bbury=480pt,%
  height=65mm,width=80mm,angle=0]{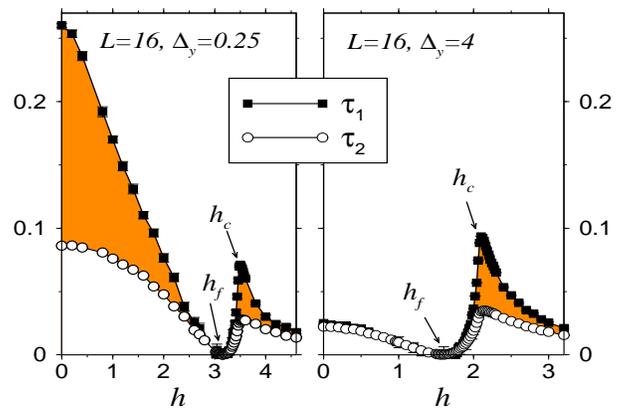}
  \null \vskip -1cm
 \caption{\label{fig1} One-tangle $\tau_1$ and sum of squared
 concurrences $\tau_2$ as a function of the applied field 
 for the 2D $S{=}1/2$ XYX model with 
 $\Delta_y {=} 0.25$ and $\Delta_y{=}4$. The vanishing 
 of $\tau_1$ signals the occurrence of an exactly factorized
 state, while its spike signals the quantum critical point.}
 \end{center}
% \null\vskip -.8cm 
\end{figure}

In Fig.~\ref{fig1} we plot $\tau_1$ and $\tau_2$ for the 
2D XYX model for two values of $\Delta_y$
in the EP and EA case. The most striking
feature is the non-monotonic behavior of both
quantities as a function of the field. Although the field
is expected to suppress quantum fluctuations and entanglement
only in the extreme $h{\to}\infty$ limit, we observe that 
there exists an intermediate non-trivial value $h_{\rm f}$ 
at which entanglement disappears completely, since $\tau_1$
is vanishing. This signals an 
\emph{exactly factorized ground state} in the 2D model, completely
unknown before. Above the factorizing field $h_{\rm f}$ 
both $\tau_1$ and $\tau_2$ have a steep recovery, that 
will later be associated with the occurrence of the quantum
phase transition. Finally, we observe that the CKW conjecture
is verified for any field value within error bars.

The behavior of the concurrence is 
found to be qualitatively the same as that observed~\cite{Roscildeetal04} 
in D=1 and we do not report it here;
however we underline that also in D=2 its range stays extremely 
short  and that, for all values of $\Delta_y$ studied, 
$C_{ij}^{(1)}$ and $C_{ij}^{(2)}$ cross each other 
(and vanish simultaneously) at the factorizing field. 
In particular, for $h{>}h_{\rm f}$, $C_{ij}^{(1)} {>} C_{ij}^{(2)}$, 
so that Eqs.~(\ref{e.C1})-(\ref{e.C2})
are accurate in the most interesting part of the 
ordered phase, namely above the factorizing field
and all the way to the critical point.

The existence of a factorized ground state in the 1D XYX model
has been exactly proven in Ref.~\onlinecite{Kurmannetal82},
and its entanglement signature has been studied in 
Ref.~\onlinecite{Roscildeetal04}. 
The evidence for a factorized ground state in D=2,
coming from our QMC simulations, 
lead us to the 2D generalization of the exact 
proof of factorization: We were indeed 
able to demonstrate that a factorized ground state exists for the
most general Hamiltonian Eq.~(\ref{e.XYZhz})
on any 2D bipartite lattice. The proof will be soon
reported elsewhere~\cite{Verrucchietal05}, but we here outline
the essential findings.

For any value of the anisotropies $\Delta_y$ and $\Delta_z$, there exists an
ellypsoid in field space
\begin{eqnarray} 
  \frac{h_{x}^2}{(1{+}\Delta_y)(1{+}\Delta_z)}
   &{+}&\frac{h_{y}^2}{(1{+}\Delta_y)(\Delta_y{+}\Delta_z)}{+}\nonumber\\
   &{+}&\frac{h_{z}^2}{(1{+}\Delta_z)(\Delta_y{+}\Delta_z)}=4 
 \label{e.hfact}
\end{eqnarray} 
such that, when $\bm h$ lies on its surface, the ground state of
the corresponding model is factorized, $|\Psi\rangle {=} \bigotimes_{i=1}^{N}
|\psi_i\rangle $. The single-spin states $|\psi_i\rangle$ are eigenstates of
$({\bm n}_I \cdot \hat{\bm S})$, ${\bm n}_I$ being the local spin orientation on
sublattice $I$. We will hereafter indicate with ${\bm h}_{\rm f}$ ({\it
factorizing} field) the field satisfying Eq.~(\ref{e.hfact}); at ${\bm h}{=}{\bm
h}_{\rm f}$, the reduced energy per site is found to be
$\epsilon{=}{-}(1{+}\Delta_y{+}\Delta_z)/2$. In
the particular case of $\Delta_z {=} 1$ and $\bm h {=} (0,0,h)$, the factorizing
field takes the simple expression $h_{\rm f} {=} 2\sqrt{2(1{+}\Delta_y)}$.  As for
the structure of the ground state, taking ${\bm n}_I{=}(\cos \phi_I
\sin\theta_I,\sin \phi_I\sin\theta_I,\cos \theta_I)$, the analytical
expressions for $\phi_I$ and $\theta_I$ are available via the solution of a system
of linear equations.
The local spin orientation turns out to be different in the EP and EA cases, being
$\phi_1 {=} 0$, $\phi_2 {=} \pi$, 	
 $\theta_1 {=} \theta_2 {=} \cos^{-1}\sqrt{(1{+}\Delta_y)/2}$ 
for $\Delta_y{<}1$, and
$\phi_1 {=}\pi/2$, $\phi_2 {=} {-}\pi/2$, 
 $\theta_1 {=} \theta_2 {=} 
\cos^{-1}\sqrt{2/(1{+}\Delta_y)}$ 
for $\Delta_y{>}1$.
\begin{figure}
 \begin{center}
\includegraphics[bbllx=32pt,bblly=19pt,bburx=582pt,bbury=468pt,%
   height=60mm,width=77mm,angle=0]{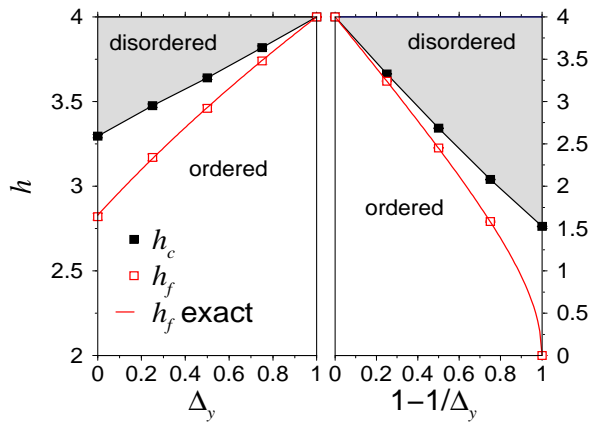}
\caption{\label{fig2} Phase diagram of the 2D XYX model
 in a field in the easy-plane case ($\Delta_y{<}1$, left panel)
 and in the easy-axis case ($\Delta_y{>}1$, right panel).
 The squares are the QMC estimates for $h_{\rm f}$ and $h_{\rm c}$.}
 \end{center} 
% \null\vskip -.8cm 
\end{figure} 
 The numerical and analytical findings for the location 
of the factorizing field in the 2D XYX model as a function 
of the anisotropy $\Delta_y$ are summarized in Fig.~\ref{fig2}.
In particular, the QMC data for $h_{\rm f}$ are obtained
from the vanishing point of the nearest neighbor concurrence,
which gives the best resolution of the location of the 
factorized state.
In the same figure, we also report the line of quantum-critical 
fields $h_{\rm c}$, extracted through the linear scaling 
of the spin-spin correlation length, 
$\xi^{xx}(h{=}h_{\rm c}){\sim} L$.

The critical scaling of the structure factor 
$S_{xx}(q {=} 0){\sim}L^{\gamma/\nu{-}z}$ at the 
critical field is fully consistent with 
the best estimates for the critical exponents of the 
2D Ising model in a transverse field, $z{=}1$ 
$\gamma{=}1.237$, and 
$\nu{=}0.630$~\cite{PelissettoV02,Chakrabartietal96}. 
We observe that at the Heisenberg point $\Delta_y{=}1$ the critical field
and the factorizing field converge to the (non-critical)
saturation field $h{=}4$. 

It is not clear yet, not even in the 1D case,
if the occurrence of a factorized ground 
state be related  with that of the quantum phase transition in 
these models. In this respect we notice that the change 
in the analytical expression of the concurrence 
from $C_{ij}^{(2)}$ to $C_{ij}^{(1)}$ at $h{=}h_{\rm f}$ 
can be related to a qualitative change in the 
wavefunction~\cite{Verrucchietal05} at the level
of the phase coherence between any two spins.
The factorized ground state could hence represent a crucial
step towards a global rearrangement of the 
ground state, {\it in view} of the quantum phase transition.

\begin{figure}
 \begin{center}
\includegraphics[bbllx=0pt,bblly=19pt,bburx=600pt,bbury=460pt,%
     height=60mm,width=95mm,angle=0]{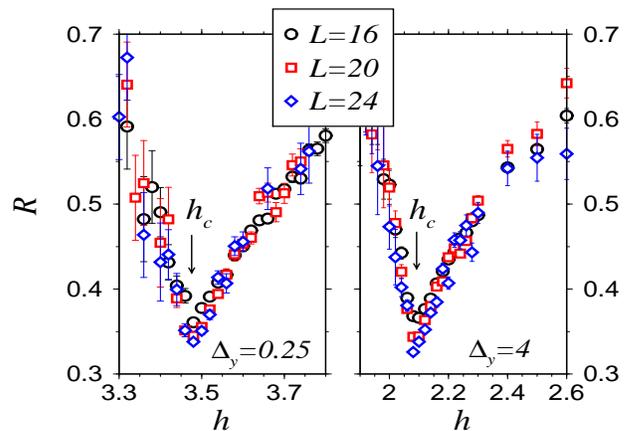}
 \caption{\label{fig3} Entanglement ratio $R=\tau_2/\tau_1$ 
 around the quantum critical point of the 2D XYX model 
 ($\Delta_y=0.25$ and $\Delta_y=4$).}
\end{center}
%\null\vskip -.8cm 
\end{figure} 
Let us now move to the analysis of the critical behavior 
of entanglement from the point of view of the entanglement ratio 
$R$ which, under the validity of the CKW conjecture, 
provides a new insight in the relative weights of multipartite 
\emph{vs} bipartite entanglement. 
Fig.~\ref{fig3} shows $R$ as a function of the field
for both the EP ($\Delta_y{=}0.25$) and the
EA ($\Delta{=}4$) case. We observe that a pronounced
dip, in the form of a cusp, is exhibited at the critical point, 
signaling a quantum-critical enhancement
of multi-spin entanglement involving $n$ spins with $n{>}2$
at the expenses of pairwise entanglement.

The critical features of $R$ in D=2 are surprisingly
analogous to those exhibited in the 1D case by the 
same class of models~\cite{Roscildeetal04}.
Such observation indicates a universal scenario for
entanglement at a quantum phase transition. We specialize
here to the case of \emph{second-order} quantum phase
transitions, assuming that the correct 
estimator of concurrence is provided by 
Eqs.~(\ref{e.C1})-(\ref{e.C2}), which implies
that, if not vanishing, 
$C_{ij}{=}g_{ij}^{zz}{-}\frac{1}{4}{+}|g_{ij}^{xx}{-}g_{ij}^{yy}|$.
As $\tau_1$ and $\tau_2$
are continuous (and non-vanishing) functions around a
quantum phase transition, $R$ is continuous as well, and 
its critical behavior is best predicted
from its derivative with respect to the dimensionless
control parameter, hereafter indicated as
$\lambda$ ($\lambda{=}h$ in the transition under 
investigation).

At the quantum phase transition one of the
magnetizations, \emph{e.g.} $M^{x}$, will vanish,
%$M^{x}\frac{\sim}{\lambda{\to}\lambda_{\rm c}^-} 
%(\lambda_{\rm c}{-}\lambda)^{\beta}$. 
$M^{x} \sim (\lambda_{\rm c}{-}\lambda)^{\beta}$
for $\lambda{\to}\lambda_{\rm c}^-$.
Therefore the $\lambda$-derivative of $\tau_1$ for 
$\lambda{<}\lambda_{\rm c}$ contains the term
${-}\partial_{\lambda}(M^{x})^2{\sim}(\lambda_{\rm c}{-}\lambda)^{2\beta{-}1}$,
which diverges to ${+}\infty$ if $\beta{<}1/2$
(\emph{i.e.}, away from the mean-field limit).
In presence of an applied magnetic field
along, \emph{e.g.} the $z$ axis, the derivative 
${-}\partial_{\lambda} (M^{z})^2$ can also be divergent, 
this time to ${-}\infty$. This is indeed the case for the 
1D transverse Ising model~\cite{Niemeijer67} where ${-}\partial_{\lambda} 
(M^{z})^2{\sim}{-}\ln(\lambda{-}\lambda_{\rm c})$. 
On the other hand, in the mean-field 
(D=$\infty$) limit
no divergence of the derivative is present, but only a 
discontinuity jump~\cite{Pfeuty70}. Therefore we expect
the singularity in the derivative to become weaker when
the dimensionality is increased, as our 2D results
suggest when compared with the 1D results of 
Ref.~\cite{Roscildeetal04}. This means that, within
the Ising universality class and below the upper
critical dimension, $\partial_\lambda \tau_1$ is dominated
by the power-law divergence to ${+}\infty$ of 
${-}\partial_{\lambda} (M^{x})^2$. This is 
\emph{a fortiori} true in the case of phase transitions
with $\beta{<}1/2$ in which the only ``singular'' magnetization
is the order parameter.
For $\lambda{\to}\lambda_{\rm c}^+$, instead, $\partial_{\lambda}\tau_1$ 
is either finite or, again in presence of a field
along the $z$ axis, it might be dominated by
the divergence to ${-}\infty$ of ${-}\partial_{\lambda} (M^{z})^2$.
Therefore $\tau_1$ has a discontinuous derivative at $\lambda_{\rm c}$,
and, being a continuous function, it shows a \emph{cusp maximum}
at the critical point. This feature witnesses the 
quantum critical enhancement of global entanglement, 
as estimated through $\tau_1$.

 On the other hand, $\tau_2$ is essentially a sum of spin-spin
correlators $g^{\alpha\alpha}_{ij}$ with $|i{-}j|$ limited
by the very short range of the 
concurrence~\cite{Osterlohetal02,Roscildeetal04}. It is fair to 
assume that $\tau_2$ is dominated by the nearest-neighbor (n.n.)
concurrence, which contains the most
relevant features for this analysis. In particular, 
the nearest-neighbor correlators building up the 
n.n. concurrence are also the fundamental ingredients
of the energy, along with the magnetization $M^z$ 
in presence of a field. At a $T{=}0$ continuous transition 
the first derivative of the energy w.r.t. $\lambda$ is 
continuous, so that either the derivative of the 
correlators is not-singular, or, if
a singularity shows up in $\partial_{\lambda} M^{z}$,
it has to be compensated by an equal and opposite
singularity in one of the n.n. correlators. This suggests
that $\partial_{\lambda}\tau_2$ is
at most as singular as $\partial_{\lambda} M^{z}$,
with the same sign of the possible singularity
~\cite{Osterlohetal02,OsborneN02}.

 Finally, we consider 
 $\partial_\lambda R {=} (\partial_{\lambda}\tau_2)/\tau_1{-}
 \tau_2(\partial_\lambda\tau_1)/(\tau_1)^2$. For 
$\lambda{\to}\lambda_{\rm c}^-$, if $\beta{<}1/2$ and under the 
conditions of $\partial_{\lambda} M^{z}$ having a weaker singularity
than $\partial_{\lambda} (M^{x})^2$, 
$\partial_\lambda R$ is dominated by ${-}\partial_\lambda \tau_1$
with a power-law divergence to ${-}\infty$. 
On the other hand, for $\lambda{\to}\lambda_{\rm c}^+$, 
$\partial_\lambda R$ might be non-singular or, at most,
as singular as $\partial_{\lambda} M^{z}$, with a 
divergence to ${+}\infty$. We therefore obtain that
$R$ has derivatives of opposite 
sign when approaching the critical point from left
and right, and, being $R$ a continuous function, 
$\lambda{=}\lambda_{\rm c}$ can only be a \emph{cusp minimum}.

In conclusion, through a systematic study 
of the entanglement of formation, we have shown that 
anisotropic $S{=}1/2$ quantum
Heisenberg antiferromagnets on the square lattice 
in an arbitrarily oriented
uniform magnetic field display an exactly
factorized state for a given field intensity, depending
on the orientation of the field and on the anisotropies.
The existence of many real compounds whose magnetic behaviour 
is described by the Hamiltonian Eq. (\ref{e.XYZhz})
suggests the possibility of an experimental analysis of 
our findings. Indeed, the existence of a classical-like ground 
state with flat correlators and absence of quantum fluctuations
should lead to striking signatures both in the static and
dynamical observables.

Finally, we have also shown that the field-induced quantum phase 
transition occurring in these systems is characterized by critical 
enhancement of \emph{multipartite} entanglement, 
in the form of a cusp minimum in the pairwise-to-global 
entanglement ratio $R$. 
Based on general scaling arguments, we conclude that such a
cusp in $R$ is a universal entanglement signature of
quantum critical points for continuous quantum phase 
transitions below the upper critical dimension.

This work has been supported 
by NSF under grant DMR-0089882 (T.R. and S.H.),
by INFN, INFM, and MIUR-COFIN2002 (A.F., P.V., and V.T.).

\end{document}